\documentclass[conference]{IEEEtran}

\usepackage{marvosym}
\usepackage{amsmath}
\usepackage{amssymb}
\usepackage{booktabs}
\usepackage{tabularx}
\usepackage[margin=2cm]{geometry}
\usepackage{tikz}
\usepackage{graphicx}
\usepackage{color}
\usepackage[T1]{fontenc}
\usetikzlibrary{calc}
\usepackage{tikz}
\usetikzlibrary{calc}
\usetikzlibrary{shapes.geometric, arrows}
\tikzstyle{startstop} = [ellipse, text centered, draw=black, semithick]
\tikzstyle{io} = [trapezium, trapezium left angle=70, trapezium right angle=110, minimum width=0.3cm, minimum height=0.3cm, text centered, draw=black, semithick]
\tikzstyle{process} = [rectangle, minimum width=1cm, minimum height=0cm, text centered,    draw=black, semithick]
\tikzstyle{decision} = [diamond, minimum width=1cm, minimum height=0.5cm, text centered,  draw=black, semithick]
\tikzstyle{arrow} = [semithick,->,>=stealth]
\usepackage{subfig}
\usepackage{siunitx}
\usepackage{verbatim}

\usepackage[ruled, vlined,linesnumbered]{algorithm2e}

\usetikzlibrary{decorations.pathreplacing}

\tikzset{radiation/.style={{decorate,decoration={expanding waves,angle=90,segment length=4pt}}},
         relay/.pic={
        code={\tikzset{scale=5/10}
            \draw[semithick] (0,0) -- (1,4);
            \draw[semithick] (3,0) -- (2,4);
            \draw[semithick] (0,0) arc (180:0:1.5 and -0.5) node[above, midway]{#1};
            \node[inner sep=4pt] (circ) at (1.5,5.5) {};
            \draw[semithick] (1.5,5.5) circle(8pt);
            \draw[semithick] (1.5,5.5cm-8pt) -- (1.5,4);
            \draw[semithick] (1.5,4) ellipse (0.5 and 0.166);
            \draw[semithick,radiation,decoration={angle=45}] (1.5cm+8pt,5.5) -- +(0:2);
            \draw[semithick,radiation,decoration={angle=45}] (1.5cm-8pt,5.5) -- +(180:2);
  }}
}

\tikzset{radiation/.style={{decorate,decoration={expanding waves,angle=90,segment length=4pt}}},
         relay1/.pic={
        code={\tikzset{scale=7/10}

    \draw[thin] (0,0) -- (0,10);
    \draw[thin] (5,0) -- (5,10);
    \draw[thin] (0,0) -- (5,0);
    \draw[thin] (0,10) -- (5,10);

    \draw[thin] (0.75,1.5) -- (0.75,8.5);
    \draw[thin] (0.75,8.5) -- (4.25,8.5);
   \draw[thin] (4.25,8.5) -- (4.25,1.5);
   \draw[thin] (4.25,1.5) -- (0.75,1.5);

   \draw[thin] (2.5,0.7) circle(15pt);
   \draw[thin] (2.5,9.5) circle(10pt);

       \draw[thin] (1.5,9) -- (3.5,9);
       \draw[thin] (1.5,8.7) -- (3.5,8.7);
       \draw[thin] (1.5,9) -- (1.5,8.7);
       \draw[thin] (3.5,9) -- (3.5,8.7);

  }}
}

\usetikzlibrary{arrows.meta,calc,decorations.markings,math,arrows.meta}
\usetikzlibrary{shapes.multipart}

\usepackage{array}
\usepackage{makecell}
\newcolumntype{x}[1]{>{\centering\arraybackslash}p{#1}}

\usepackage{tikz}
\usetikzlibrary{calc}
\usepackage{zref-savepos}

\newcounter{DiagonalizedEntry}
\renewcommand*{\theDiagonalizedEntry}{NTE-\the\value{DiagonalizedEntry}}

\ifCLASSINFOpdf
\else
\fi

\begin{document}

\title{Rate-Splitting Random Access Mechanism for Massive Machine Type Communications in 5G Cellular Internet-of-Things}


\author{\IEEEauthorblockN{Yeduri Sreenivasa Reddy$^{1}$, Garima Chopra$^{2}$, Ankit Dubey$^{3}$, Abhinav Kumar$^{2}$, Trilochan Panigrahi$^{1}$, \\ and Linga Reddy Cenkeramaddi$^{4}$}
	\IEEEauthorblockA{
		$^{1}$Department of ECE, National Institute of Technology Goa, Ponda, Goa 403401, India \\
	$^{2}$Department of EE, Indian Institute of Technology Hyderabad, Kandi, Sangareddy, Telangana 502285, India \\ 
		$^{3}$Department of EE, Indian Institute of Technology Jammu, Jammu \& Kashmir 181221, India \\
		$^{4}$Department of ICT, University of Agder, Grimstad-4879, Norway\\
		Email: $^{1}$\{ysreenivasareddy,tpanigrahi\}@nitgoa.ac.in, $^{2}$\{garima.chopra,abhinavkumar\}@ee.iith.ac.in,
		\\ $^{3}$ankit.dubey@iitjammu.ac.in, $^{4}$linga.cenkeramaddi@uia.no 
		}} 

\maketitle
 \thispagestyle{empty}
 \pagestyle{empty}
\pagestyle{plain} 
\begin{abstract}
The cellular Internet-of-Things has resulted in the deployment of millions of machine-type communication (MTC) devices. These massive number of devices must communicate with a single gNodeB (gNB) via the random access channel (RACH) mechanism. However, existing RACH mechanisms are inefficient when dealing with such large number of devices. To address this issue, we propose the rate-splitting random access (RSRA) mechanism, which uses rate splitting and decoding in rate-splitting multiple access (RSMA) to improve RACH success rates. The proposed mechanism divides the message into common and private messages and enhances the decoding performance. We demonstrate, using extensive simulations, that the proposed RSRA mechanism significantly improves the success rate of MTC in cellular IoT networks. We also
evaluate the performance of the proposed mechanism with increasing number of devices and received power difference.
\end{abstract}
\par \textbf{\textit{Keywords}--
Cellular Internet-of-Things, gNodeB, machine type communications (MTC), random access channel, rate-splitting multiple access, received power.}

%
\IEEEpeerreviewmaketitle

\section{INTRODUCTION}

The cellular Internet-of-Things (C-IoT) has transformed wireless communications from human-centric to human-independent. According to an IHS Markit report \cite{ihs-report}, this will result in the deployment of approximately 125 billion smart devices by 2030. Human-centric communication is also called as human-to-human communication, and it is primarily intended for voice or video calls, web browsing etc. \cite{taleb-cm12}. Machine type communications (MTC), on the other hand, which is human independent communication, is designed to support millions of devices, energy efficiency, and so on \cite{3gpp-ser}.

\begin{figure}[!t]
\centering
\includegraphics[height= 2.5 in, width=3.45 in]{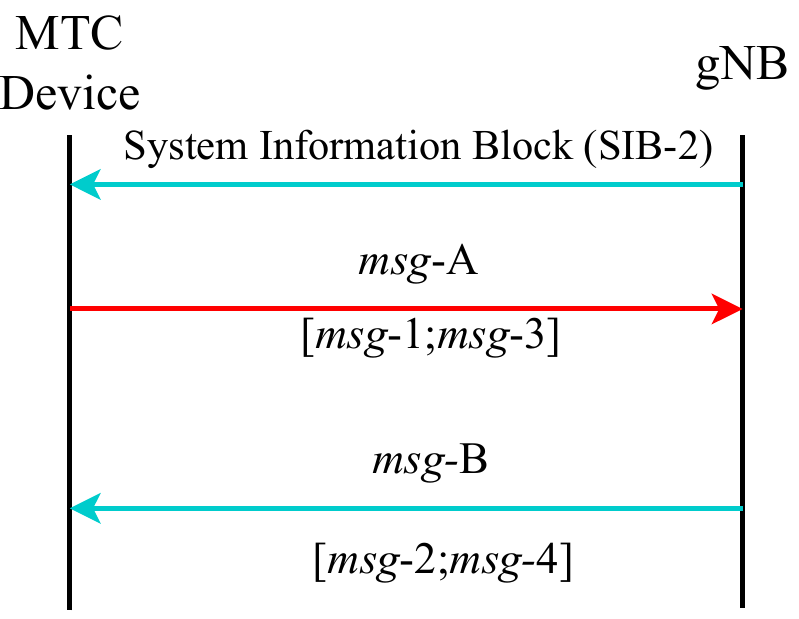}
\caption{RACH message exchange for 5G.}
\label{rach_5g}
\end{figure}

Since the MTC devices have to be energy efficient, each device wakes up to transmit the collected data and then returns to sleep. However, in order for information to be exchanged, the device must be connected to a gNodeB (gNB). A random access channel (RACH) mechanism is the mechanism by which a device transitions from a radio resource control (RRC)-idle state to an RRC-connected state. 3rd generation partnership project (3GPP) has proposed an long-term evolution (LTE) RACH mechanism based on four message exchanges \cite{3gpp3}. To reduce the latency and control-signaling overhead of LTE RACH, 3GPP has proposed a new RACH mechanism for 5G known as early data transmission (EDT), in which there is only two message exchange between each device and gNB, as shown in Fig. \ref{rach_5g} \cite{3gpp5}. 

In EDT, similar to the LTE RACH, the gNB broadcasts system information block (SIB)-2 message to all the devices under its coverage. All the devices with a packet to transmit reads the SIB-2. Here, SIB-2 indicates the preamble group to be used and configuration of preamble to resources for \textit{msg}-3. After successful decoding of SIB-2, the device starts the RACH mechanism by transmitting message (\textit{\textit{msg}})-A, which is a combination of
preamble \textit{\textit{msg}}-1 and RRC connection request (data or device identity) \textit{\textit{msg}}-3 to the gNB. After successful reception of \textit{\textit{msg}}-A, gNB responds with \textit{\textit{msg}}-B, which is the combination of RA message \textit{\textit{msg}}-2 and RRC Contention resolution \textit{\textit{msg}}-4. The messages from \textit{\textit{msg}}-1 to \textit{\textit{msg}}-4 are same as defined for LTE RACH. There are three types of responses possible from the gNB to an MTC device based on the type of message received. The first one is no response due to the failure in receiving \textit{\textit{msg}}-A. In this case, the device takes a backoff and starts either two message exchange or falls back to four message exchange of LTE. The second one is the successful reception of \textit{\textit{msg}}-1 but, failure to decode \textit{\textit{msg}}-3. In this case, the gNB responds with \textit{\textit{msg}}-2 (RA response) of four message exchange to indicate the PUSCH for retransmission of \textit{\textit{msg}}-3 of \textit{\textit{msg}}-A. Then, the gNB responds with \textit{\textit{msg}}-4 of \textit{\textit{msg}}-B. The third one is the successful reception of both the messages of \textit{\textit{msg}}-A. In this case, the gNB responds with \textit{\textit{msg}}-B (\textit{msg}-2+\textit{msg}-4) that indicates the contention resolution ID. Successful reception of \textit{\textit{msg}}-B indicates successful RACH \cite{3gpp5}. 

The two-phase cluster-based group paging scheme was proposed in \cite{a1} to handle massive MTCs simultaneous channel access and overcome the limitations of the 3GPP. As a result of the large number of connection requests, congestion is a major limiting factor. Extended access barring (EAB) has been proposed as a connection control mechanism in 3GPP specifications. The EAB has been studied in \cite{a2} and two improved EAB schemes have been proposed. Similarly, \cite{a3} has proposed a distribution method based on non-orthogonal multiple access (NOMA) and  Q-learning to dynamically allocate random access slots and handle massive MTC networks. The orthogonal sequences generated from both Gaussian distribution and the Zadoff-Chu (ZC) have been studied and analyzed for asymptotic behavior for success probabilities using NOMA, and the closed-form expression has been derived in \cite{a5}. A Markov chain-based access barring (M-ACB) to reduce congestion of both delay-tolerant devices and delay-sensitive devices to properly utilize network resources has been studied in \cite{a4}. A distributive approach has been proposed in \cite{a6} to estimate the optimal back-off parameters according to precise interpretations of the observed statistics. With such a large number of MTC devices, resource allocation complicates the concern even more. To manage resources among MTC devices, a scheme has been proposed in \cite{a7} to dynamically prioritise the MTC devices.

In \cite{liang-twc17}, the authors have proposed a non-orthogonal random access (NORA) mechanism that utilizes intra-slot SIC to decode the \textit{\textit{msg}}-3 of collided devices. In this mechanism, the gNB can successfully decode the \textit{\textit{msg}}-3 of two collided devices if the time in their \textit{\textit{msg}}-3 arrival is greater than a threshold. A successive interference cancellation (SIC)-based RACH mechanism has been proposed in \cite{reddy-tvt21} that utilizes repetition at the device side and inter-slot SIC at the gNB. Moreover, a SIC-NORA mechanism has been proposed in \cite{reddy-access21} that utilizes both intra-slot and inter-slot SIC at gNB to further enhance the number of successful devices. However, the mechanisms in \cite{reddy-tvt21,reddy-access21} increases the network congestion and energy consumption due to the repetitions of \textit{\textit{msg}}-3.

Recently, RSMA has gained a lot of attention due to its robust transmission and general framework in comparison to NOMA and space division multiple access (SDMA). It has been shown in \cite{r2} that RSMA significantly improves the achievable data rate in comparison to NOMA and SDMA. Several existing works focus on downlink and uplink transmission \cite{r2,r4,r5,r6,r1,zhu-wcsp17}. However, RSMA can improve the quality of service and spectral efficiency for downlink transmission under the perfect channel state information of transmitter (CSIT) \cite{r1,r3} or imperfect CSIT. In RSMA, each user's message to be transmitted is split into two sub-messages such as the common and private message. The sub-messages of different users are then super positioned and transmitted to the receiver. Further, the receiver decodes the respective sub-messages by applying SIC. However, RSMA faces few challenges during implementation \cite{r3}. Firstly, the decoding order of sub-messages needs to be carefully designed. Secondly, resource management for transmission of messages. Several works study the downlink RSMA. However, only a few existing works concentrate on uplink RSMA for performance improvement \cite{zhu-wcsp17}. Further, the RACH mechanism has not been studied with RSMA. Motivated by this, we propose rate-splitting random access (RSRA) mechanism in this work. In the proposed mechanism, the MTC devices with the same preamble use the same resource for \textit{\textit{msg}}-3 transmission. On the other hand, the gNB decodes the preamble first, and then it applies SIC to decode the rest of sub-messages from \textit{\textit{msg}}-3 of each device based on the received power. This will enhance the number of devices that are getting success in each time slot. The key contributions of this work are as follows:
\begin{itemize}
    \item We propose a novel RSMA based RACH mechanism named as the RSRA mechanism for MTC in 5G. 
    \item We present numerical results to obtain the number of contending devices and the corresponding successful devices in each time slot.
    \item We discuss the effect of increasing number of MTC devices on the performance of the proposed mechanism.
    \item We present simulation results that demonstrate the impact of the received power difference between the \textit{msg}-3 of collided devices on the performance of the proposed RSRA mechanism.
    \item We present numerical results comparing the proposed mechanism's performance to that of state-of-the-art RACH mechanisms. 
\end{itemize}

The remainder of the paper is framed as follows: 
In Section~\ref{sec_sm}, we discuss the system model and the proposed RSRA mechanism. Through extensive simulation results, the performance of the proposed mechanism is evaluated in Section~\ref{RES}. Finally, Section~\ref{CON} provides the conclusion along with possible future work.



\begin{figure*}[!t]
\centering
\includegraphics[height= 3 in, width=7 in]{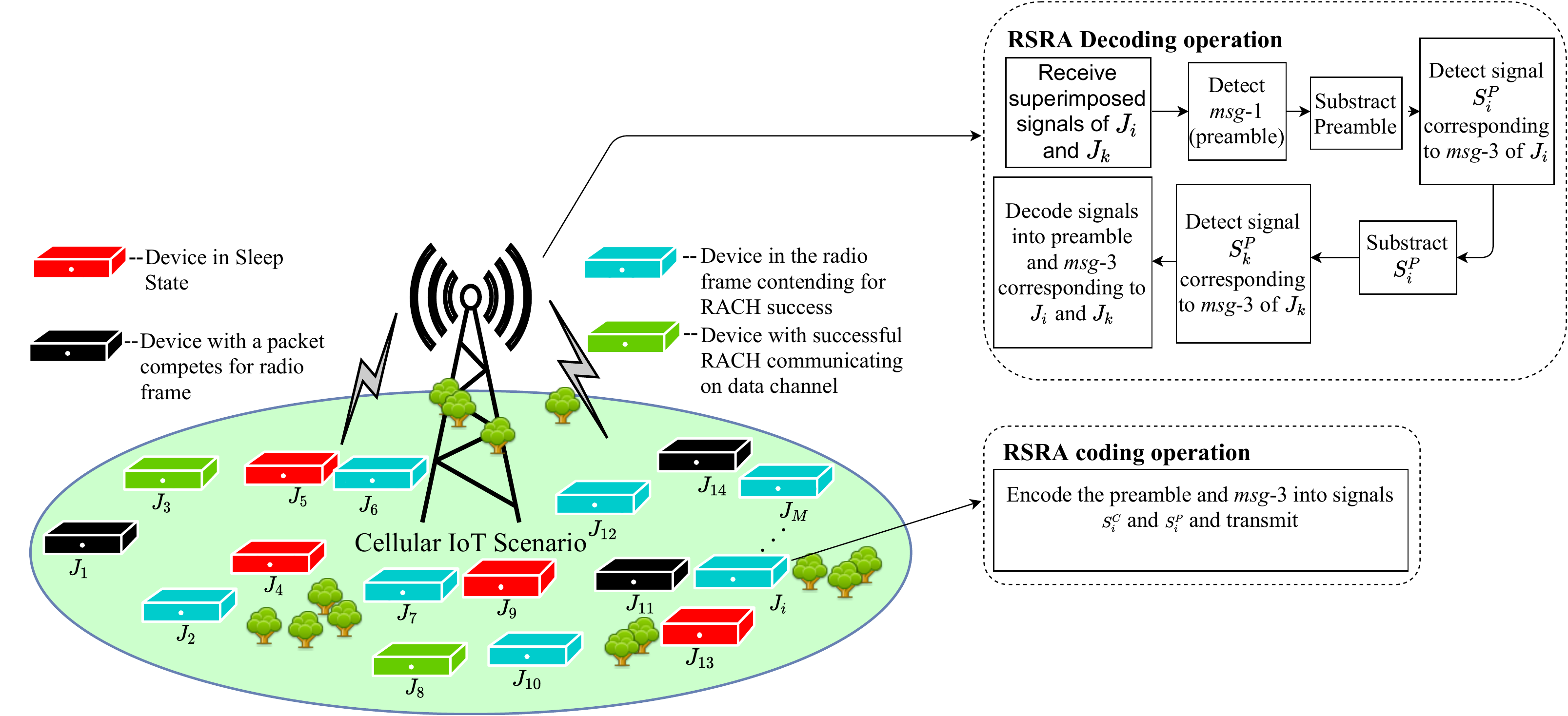}
\caption{System model.}
\label{sm}
\end{figure*}

\section{SYSTEM MODEL} \label{sec_sm}
We consider a C-IoT scenario wherein, $M$ devices namely $J_i$ $\forall $ $i \in \{1, 2, \cdots, M\} $ deployed randomly under the coverage of a single gNB $b$. These MTC devices are in either of the four states as shown in Fig. \ref{sm} \cite{reddy-tvt21}. Devices with no data to transmit remain in the sleep state. The device with a packet to transmit wake up and competes for radio frame for the initiation of RACH mechanism. The device in the radio frame competes for RACH success. All MTC devices that have successfully completed RACH transmit data using a hybrid automatic repeat request.

\begin{algorithm}[!t]
	\SetAlgoLined
	
	\KwIn{$P$}
	\KwOut{RACH success}
	\eIf{There is data\label{pkt_avl_eab}}
	{
		Uniformly generate a number $n$ from $[0,1]$ \label{gen_n_eab}\; 
		\eIf{$n<P$}
		{	
			Select a time slot from the set $\{1,2,\cdots,T\}$ uniformly at random \label{select_slot_eab}\;
			Wait for the transmission slot \label{next_slot_eab}\; 
			Select a preamble from the set $\{0,1,\cdots,K\}$ uniformly\;
			Start RACH mechanism by transmitting the preamble followed by \textit{\textit{msg}}-3\label{start_rach_eab}\;
			
			\uIf{RACH success}
			{
				Go to Step \ref{pkt_avl_eab}\;
			}
			\uElseIf{There is a collision in \textit{\textit{msg}}-3}
			{
			    Sort the $P_R$ of \textit{msg}-3 of all the collided devices in descending order\;
				\While{The difference in $P_R$ of two devices in the descending order set is greater than $\Delta P$}
				{
					RACH success with SIC\; 
					Go to Step \ref{pkt_avl_eab}\;
				}
			}
			\Else{
				Wait for the current radio frame to finish and go to Step \ref{gen_n_eab}\;
			}
			
		}		
		{
			Wait for the current radio frame to finish and go to Step \ref{gen_n_eab}\;
		}
	}
	{
		Wait for the current radio frame to finish and go to Step \ref{pkt_avl_eab}\;	
	}
	
	\caption{Proposed RSRA algorithm.}
	\label{algo-rsma}
\end{algorithm}

\subsection{RSRA coding and decoding operation}
The proposed RSRA coding and decoding operation is described in Fig. \ref{sm}.
Each MTC device transmits the encoded \textit{msg}-A consisting of a preamble and an RRC connection request as signals $s_k^c$ and $s_k^p$, respectively. In particular, $s_k^c$ and $s_k^p$ represent the common signal and private signal of the $k$th device. 
On the other end, the gNB responds with \textit{msg}-B, which is the random access response message followed by contention resolution message, after successfully decoding the \textit{msg}-A. In case only one device selects a given preamble, the gNB can successfully decodes \textit{msg}-A and responds with \textit{msg}-B to all such devices. If more than one device select a given preamble, they use the same resource to transmit \textit{msg}-3 that leads to collision in \textit{msg}-3. To all such messages, the gNB applies the RSRA decoding mechanism to successfully decode \textit{msg}-3. The decoding operation is performed in a series of steps. The RSRA decoding order for two users $J_m$ and $J_n$ is shown in Fig. \ref{sm}. Because both devices have selected and transmitted the same preamble, first part of the {\it msg}-A ($s_k^c$) can be decoded at gNB. Thereafter, gNB sequentially decodes, the second part of the {\it msg}-A ($s_K^p$) based on their received powers for $k=m$ and $k=n$. This procedure is repeated until all collided devices' \textit{msg}-3 have been successfully decoded.

\subsection{Proposed RSRA Mechanism}

Similar to 3GPP EAB mechanism, in the proposed mechanism, the gNB broadcasts SIB-2 message that indicates the configuration information of preamble and \textit{msg}-3 and access barring parameter $P$ as given in Algo. \ref{algo-rsma}. 

Each MTC device generates $n$ uniformly from $[0,1]$ upon successful reception of SIB-2. An MTC device with $n<P$ enters into a radio frame by selecting a time slot uniformly from $\{1, 2, \cdots, T\}$. A device that has selected a time slot $t$, waits for $t-1$ time slots and transmits its \textit{msg-A}, which is concatenation of a preamble followed by \textit{msg}-3 after a guard band. Here, preambles are chosen from the set $\{1, 2, \cdots, K\}$. As stated earlier, the preamble and the \textit{msg}-3 are transmitted using signals $s_k^c$ and $s_k^p$, respectively, each with a transmit power of $P_T$.

On the other end, gNB can decode the \textit{msg}-3 of a device in case it alone chooses a given preamble. Since the resources for \textit{msg}-3 transmission are configured to each preamble, the device alone transmits the \textit{msg}-3. In case a given preamble is chosen by two or more devices, they get the same resource for \textit{msg}-3 transmission. This leads to a collision. However, with RSRA mechanism, the gNB may decode the \textit{msg}-3 of such collided devices using SIC as given in Algo. \ref{algo-rsma}. The gNB starts the decoding of \textit{msg}-3 based on the difference in the powers of to \textit{msg}-3s. In case the difference of a \textit{msg}-3 of a device with highest received power and the \textit{msg}-3 of another device with second highest received power is more than $\Delta P$, the gNB can successfully decode \textit{msg}-3 of the device with highest received power \cite{zhu-wcsp17}. This process repeats until either all devices are decoded or till the difference in power levels is greater than $\Delta P$ as given in Algo. \ref{algo-rsma}. The derivation of the expression for $P_R$ is discussed in the following section.

\subsection{Received Power Calculation}
The expression of received power, $P_R$, (in dB) is given as \cite{klozar-11}
\begin{align}
P_R = P_T - PL\,,
\end{align}
where, $P_T$ (in dB) denotes the transmitted power at device and $PL$ (in dB) denotes the propagation/path loss. 
Propagation loss is defined as the amount of power loss in the channel for transmission. For the calculation of pathloss, we consider Okumura-Hata model in this paper \cite{holma-10}. The expression for pathloss in an urban environment is obtained as
\begin{align}
    PL(dB) = A_1 + A_2 \log_{10} (r),
\end{align}
where, $r$ denote the distance of the device from
the gNB. The expressions for $A_1$ and $A_2$ are obtained as
\begin{align}
    A_1 = &69.55+ 26.16 \log_{10} (f_M) - 13.82 \log_{10} (h_{g}) \nonumber \\
    &- (1.1 \log_{10}(f_M)-0.7) h_d \,,\\
    A_2 =& 44.9-6.55 \log_{10}(h_g),
\end{align}
where, $h_g$ and $h_d$ denote the gNB height and device height in m, respectively, and $f_M$ is the operating frequency in MHz. 


\begin{figure}[!t]
  \centering
  \subfloat[]{\includegraphics[width=3.45 in, height=2.5 in]{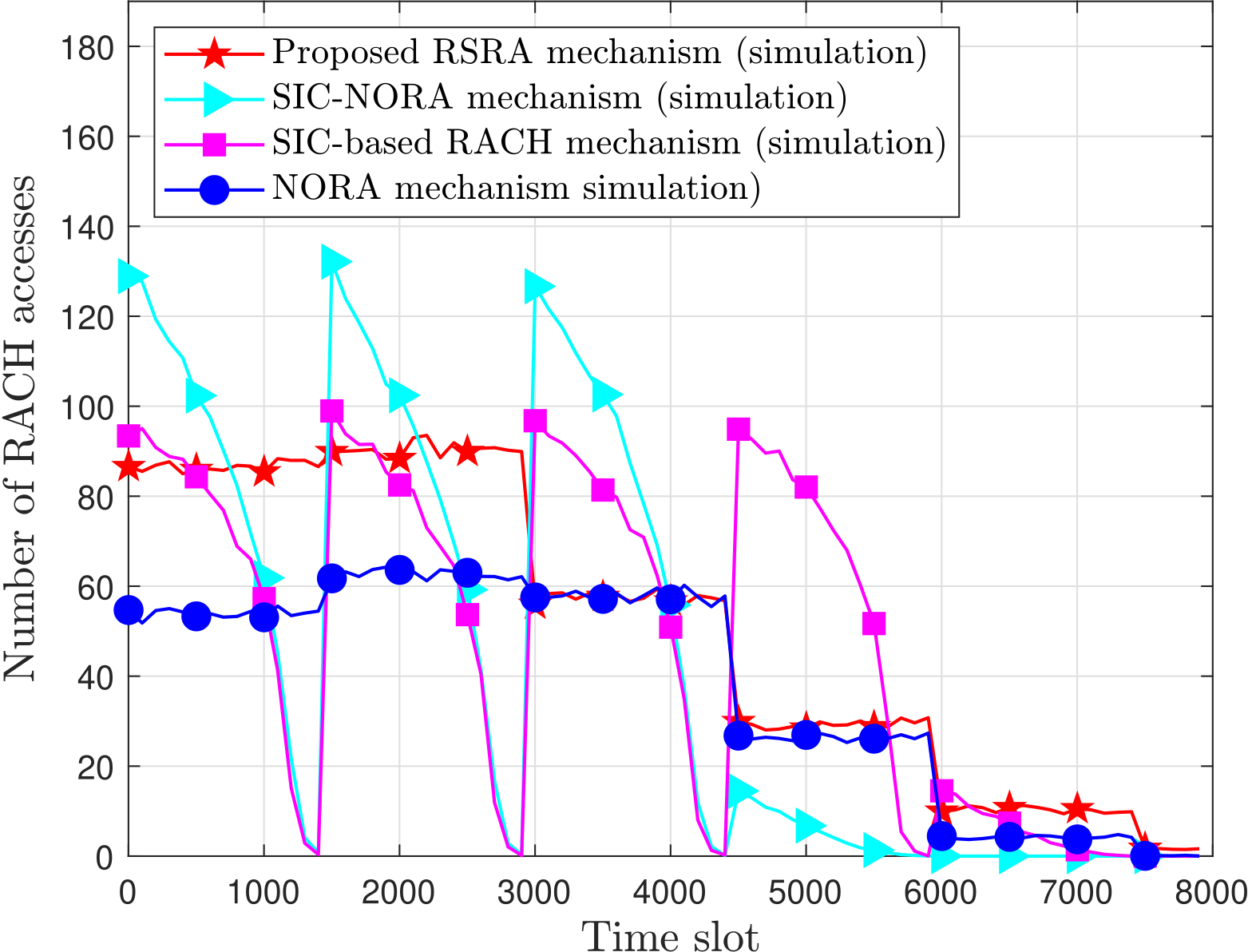}\label{num_acc}}
  \hfill
  \subfloat[]{\includegraphics[width=3.45 in, height=2.5 in]{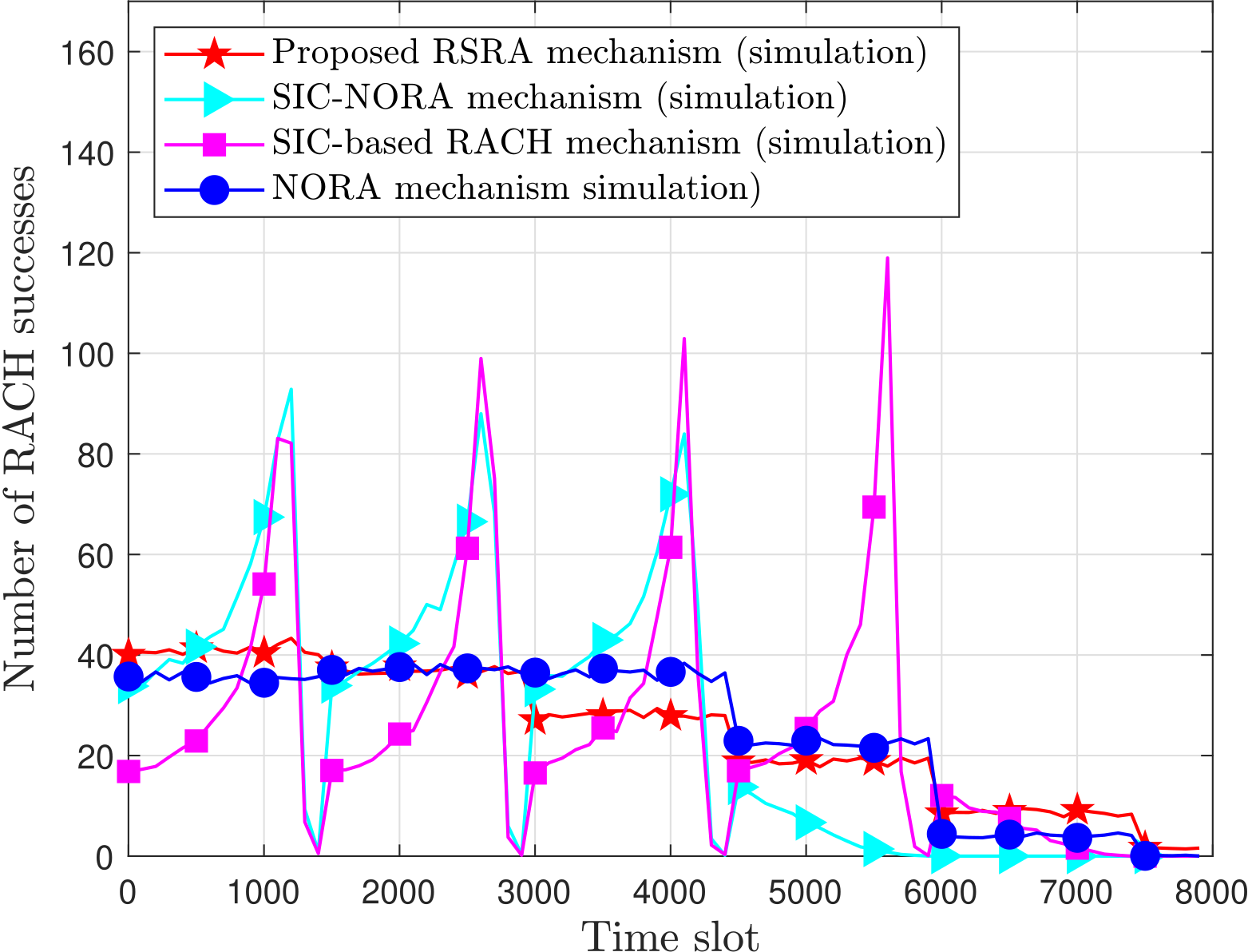}\label{num_suc}}
  \caption{The performance comparison of the proposed mechanism with the existing mechanisms in terms of (a) average number of contending devices and (b) average number of successful devices in a time slot for $T=1482$, $K=54$, $P=0.9$, and $\Delta P = 7$ dB.}
  \label{acc_suc}
\end{figure}

\section{NUMERICAL RESULTS} \label{RES}
We present the simulation results to compare the performance of the proposed mechanism with the state-of-the-art mechanisms in this section. Further, we compare the performance with increasing number of RACH accesses. Finally, we evaluate the effect of $\Delta P$.

We consider a scenario of C-IoT with $M=2\times 10^5$ MTC devices uniformly deployed in a circular area of $2$ KM radius. As per the 3GPP standards, we consider $T=1482$, $K=54$ , $P_T=24$ dBm, $f_M=1500$ MHz, $h_g = 30$ m, and $h_d=1.5$ m \cite{3gpp3, holma-10}. 

Figs. \ref{num_acc} and \ref{num_suc} show the number of contending devices and the corresponding successful devices in a time slot for all the mechanisms considered in this paper. From Fig. \ref{acc_suc}, it is observed that there is an average of $69$ devices contends in a time slot which then results in an average of $26$ successes with NORA mechanism. Further, 
there are an average of $61$ contending devices in a time slot which 
then results on an average of $33$ successes with the SIC-based RACH mechanism. Moreover, the SIC-NORA mechanism allows an average of $92$ contending devices and result in an average of $43$ successful devices. Finally, we observe that the proposed mechanism allows an average of $87$ contending devices which results in an average of $43$ successes. Even though the maximum number of RACH successes with the proposed mechanism is same as SIC-NORA mechanism, the proposed mechanism reduces the network congestion and energy consumption in comparison to SIC-NORA mechanism due to single transmission.



\begin{figure}
\centering
\includegraphics[height= 2 in, width=3.45 in]{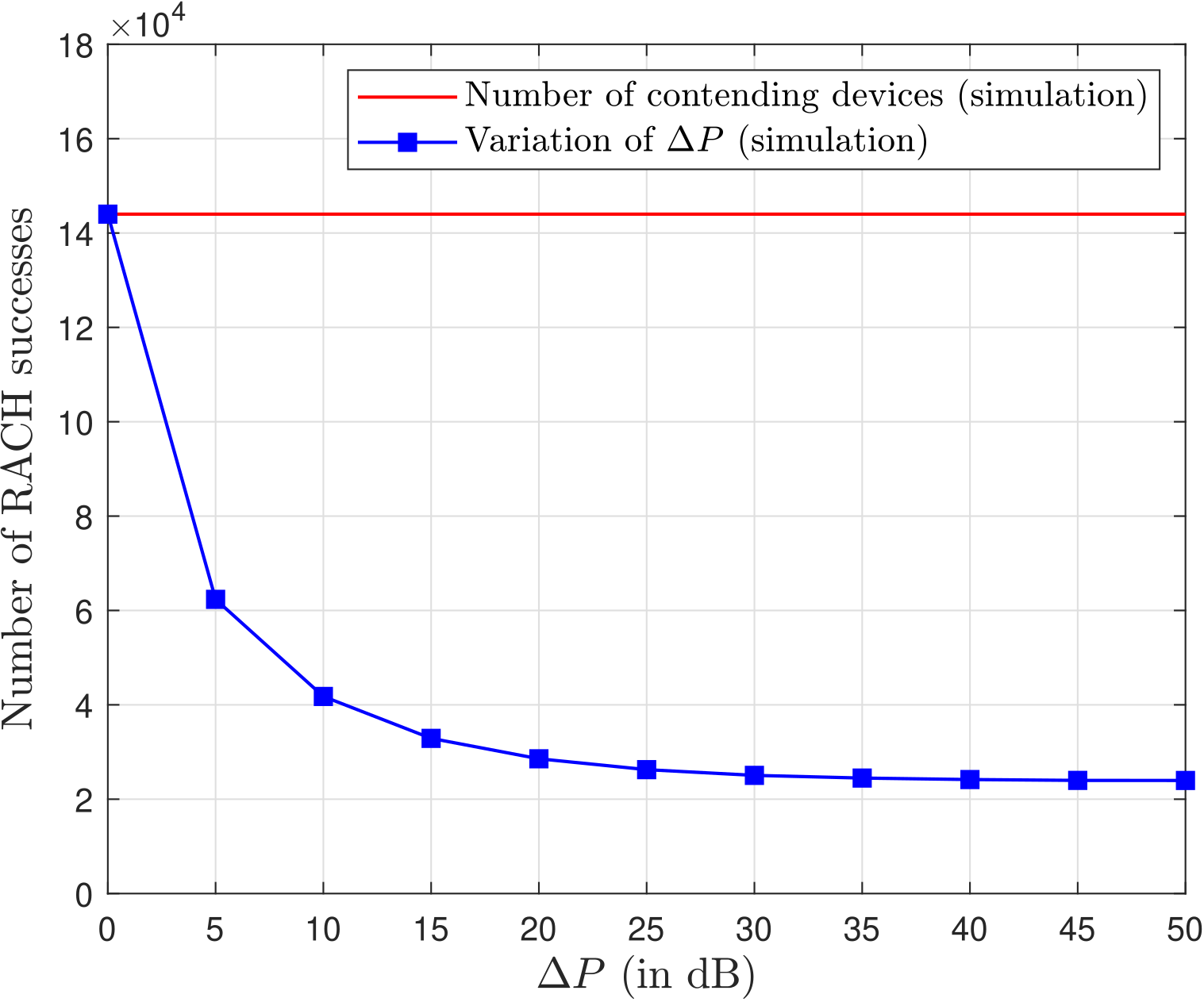}
\caption{Number of RACH successes in a radio frame vs $\Delta P$ (in dB) for the proposed RSRA mechanism with $P=1$, $T=1482$, and $k=54$.}
\label{del_P}
\end{figure}

Fig. \ref{del_P} shows the number of successful devices with increasing value of $\Delta P$ for the proposed RSRA mechanism with $1.45 \times 10^5$ contending devices. Here, we varied $\Delta P$ from $0$ dB to $50$ dB.
From Fig. \ref{del_P}, it is observed that the proposed RSRA mechanism can decode all the devices in a perfect ideal scenario, i.e. $\Delta P = 0$ dB. However, it is not possible in practice. We consider the ideal case for a fair analysis of the degradation of success rate with increasing $\Delta P$. It is observed that the number of successful devices reduces exponentially with increasing value of $\Delta P$. Thus, we conclude that the receiver design place a crucial role in enhancing the number of successful devices.  

\begin{figure}
\centering
\includegraphics[height= 2 in, width=3.45 in]{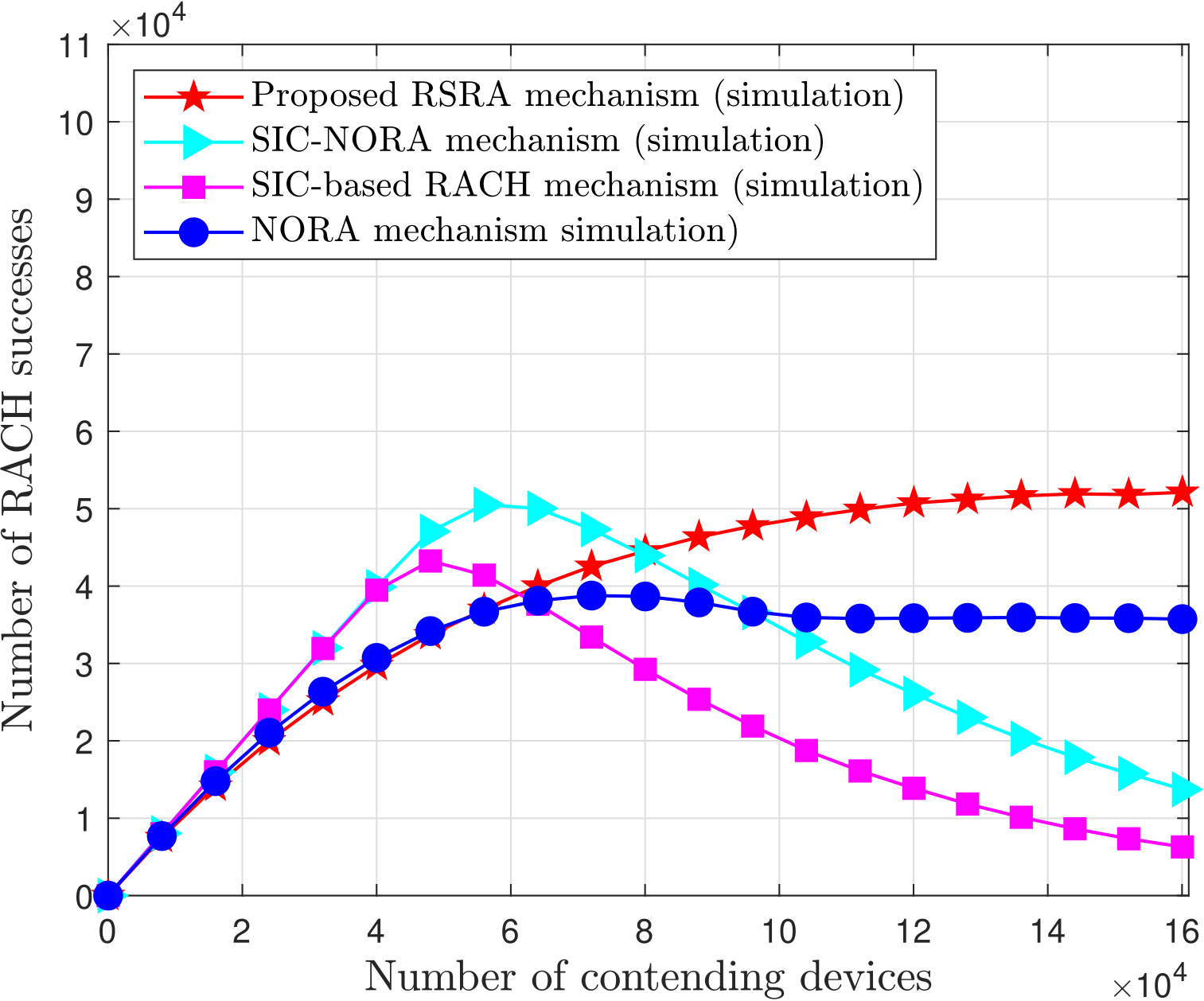}
\caption{Average number of successful devices in a radio frame with $T=1482$, $k=54$, $P=1$, and  $\Delta P = 7$ dB.}
\label{increase_dev}
\end{figure} 

Fig. \ref{increase_dev} shows the variation of number of RACH successes with increasing number of contending devices for all the mechanisms considered in this work. It is observed from Fig. \ref{increase_dev} that 
there is a maximum of $5.19 \times 10^4$ successes with the proposed mechanism. Further, it is observed that NORA mechanism, SIC-based RACH mechanism, and SIC-NORA mechanism results in a maximum of $3.82\times 10^4$, $4.19 \times 10^4$, and $5\times 10^4$ successes, respectively. 
Thus, it is concluded that the proposed RSRA mechanism outperform other mechanisms in terms of number of successful devices. Even though the performance of the proposed mechanism is comparable to SIC-NORA mechanism, the energy consumption of the proposed mechanism is less as there is no repetition rate.

\section{CONCLUSION} \label{CON}
A rate-splitting random access mechanism has been proposed in this paper to improve the number of successful devices for cellular Internet-of-Things. The MTC devices can transmit \textit{msg}-A in a randomly selected time slot under the proposed mechanism. Following that, the gNB can decode the \textit{msg}-3 of multiple collided devices with RSRA in a time slot based on their received power differences. In addition, we have evaluated the performance of the proposed mechanism with varying difference in received powers and number of contending devices. In comparison to state-of-the-art mechanisms, the proposed mechanism can support more number of devices. Furthermore, as the value of difference in received powers increases, the success rate decreases exponentially. This is due to the fact that the gNB cannot decode the devices if the received power difference is less than the threshold.
Thus, we conclude that the receiver design is critical in improving the RACH success rate.
We will derive the analytical expression for the average number of successful devices in the future for the proposed mechanism.


\begin{thebibliography}{21}


\bibitem{ihs-report}
https://cdn.ihs.com/www/pdf/IHS-Markit-Smart-Home-EBook.pdf

\bibitem{taleb-cm12}
T. Taleb and A. Kunz,
``Machine type communications in 3GPP networks: Potential, challenges, and solutions,''
\textit{IEEE Communications
Magazine,} vol. 50, no. 3, pp. 178--184, Mar. 2012.


\bibitem{3gpp-ser}
3GPP TS 22.368 v11.2.0.,
``Service requirements for machine-type
communications,'' {\em Tech. Rep. 22.368}, v16.0.0, pp. 1--32, Jul. 2020.


\bibitem{3gpp3}
3GPP, ``E-UTRA--Medium access control (MAC) protocol specification,'' {\em Tech. Rep. 36.321}, v14.3.0, pp. 1--17, Jun. 2017.

\bibitem{3gpp5}
3GPP, ``E-UTRA--Medium access control (MAC) protocol specification,'' {\em Tech. Rep. 36.321}, v16.4.0, pp. 1--17, Mar. 2021.



\bibitem{a1}
Q. Pan, X. Wen, Z. Lu, W. Jing, and L. Li, ``Cluster-based group paging for massive machine type communications under 5G networks,'' \textit{IEEE Access}, vol. 6, pp. 64891--64904, Oct. 2018.

\bibitem{a2}
J.R. Vidal, L. Tello-Oquendo, V. Pla, and L. Guijarro, ``Performance study and enhancement of access barring for massive machine-type communications,'' \textit{IEEE Access}, vol. 7, pp. 63745--63759, May 2019.

\bibitem{a3}
M.V. da Silva, R.D. Souza, H. Alves, and T. Abrão,``A NOMA-based Q-learning random access method for machine type communications,'' \textit{IEEE Wireless Communications Letters}, vol. 9, no. 10, pp. 1720--1724, Jun. 2020.

\bibitem{a5}
J. Ding, D. Qu, and J. Choi, ``Analysis of non-orthogonal sequences for grant-free RA with massive MIMO,''
\textit{IEEE Transactions on Communications}, vol. 68, no. 1, pp. 150--160, Oct. 2019.

\bibitem{a4}
J. Liu and L. Song, ``A novel congestion reduction scheme for massive machine-to-machine communication,''
\textit{IEEE Access}, vol. 5, pp. 18765--18777, Sep. 2017.

\bibitem{a6}
C. Zhang, X. Sun, J. Zhang, X. Wang, S. Jin, and H. Zhu, ``Distributive throughput optimization for massive random access of M2M communications in LTE networks,'' 
\textit{IEEE Transactions on Vehicular Technology}, vol. 69, no. 10, pp. 11828--11840, Oct. 2020. 

\bibitem{a7}

W. U. Rehman, T. Salam, A. Almogren, K. Haseeb, I. Ud Din, and S. H. Bouk, 
``Improved resource allocation in 5G MTC networks,''
\textit{IEEE Access}, vol. 8, pp. 49187--49197, Feb. 2020.

\bibitem{liang-twc17}
Y. Liang, X. Li, J. Zhang, and Z. Ding, 
``Non-orthogonal random access for 5G networks,''
\textit{IEEE Transactions on Wireless Communications,} vol. 16, no. 7, pp. 4817--4831, Jul. 2017.


\bibitem{reddy-tvt21}
Y. S. Reddy, A. Dubey, A. Kumar, and T. Panigrahi, 
``A probabilistic approach to model SIC based RACH mechanism for machine type communications in cellular networks,'' 
\textit{IEEE Transactions on Vehicular Technology,} vol. 70, no. 2, pp. 1878--1893, Feb. 2021.

\bibitem{reddy-access21}
Y. S. Reddy, A. Dubey, A. Kumar, and T. Panigrahi, 
``A successive interference cancellation based random access channel mechanism for machine-to-machine communications in cellular Internet-of-Things,''
\textit{IEEE Access}, vol. 9, pp. 8367--8380, Jan. 2021.

\bibitem{r2}
Y. Mao, B. Clerckx, and V. O. K. Li,
``Rate-splitting multiple access for downlink communication systems: bridging, generalizing, and outperforming SDMA and NOMA,'' 
\textit{EURASIP Journal on Wireless Communications and Networking}, vol. 133, May 2018.

\bibitem{r4}
J. Cao and E. M. Yeh, “Asymptotically optimal multiple-access
communication via distributed rate splitting,” IEEE Trans. Inf.
Theory, vol. 53, no. 1, pp. 304–319, Jan 2007.

\bibitem{r5}
M. Dai, B. Clerckx, D. Gesbert, and G. Caire, “A rate splitting strategy for massive MIMO with imperfect CSIT,” IEEE Transactions on
Wireless Communications, vol. 15, no. 7, pp. 4611–4624, July 2016.

\bibitem{r6}
Y. Mao, B. Clerckx, and V. O. Li, “Rate-splitting for multi-user
multi-antenna wireless information and power transfer,” in Proc.
IEEE Int. Workshop Signal Process. Advances Wireless Commun.
Cannes, France: IEEE, 2019, pp. 1–5.

\bibitem{r1}
B. Clerckx, Y. Mao, R. Schober, and H. V. Poor, 
``Rate-splitting
unifying sdma, oma, noma, and multicasting in miso broadcast
channel: A simple two-user rate analysis,'' \textit{IEEE Wireless Communications Letters}, vol. 9, no. 3, pp. 349–-353, Nov. 2020.

\bibitem{zhu-wcsp17}
Y. Zhu, X. Wang, Z. Zhang, X. Chen, and Y. Chen, 
``A rate-splitting non-orthogonal multiple access scheme for uplink transmission,''
in \textit{Proc. International Conference on Wireless Communications and Signal Processing (WCSP),} Nanjing, China, Oct. 2017, pp. 1--6.

\bibitem{r3}
B. Clerckx, H. Joudeh, C. Hao, M. Dai, and B. Rassouli, “Rate
splitting for MIMO wireless networks: A promising PHY-layer
strategy for LTE evolution,” \textit{IEEE Communications Magazine}, vol. 54, no. 5, pp. 98–-105, May 2016.




\bibitem{klozar-11}
L. Klozar and J. Prokopec, 
``Propagation path loss models for mobile communication,''
in \textit{Proc. International Conference Radioelektronika,} Brno, Czech Republic, Apr. 2011, pp. 1-4.




\bibitem{holma-10}
H.Holma and A.Toskala, 
``WCDMA for UMTS: HSPA evolution and LTE,'' 5th edition,
John Wiley \& Sons, 2010.





































\end{thebibliography}
\end{document}